\documentclass[preprint,amsmath,amssymb,aps,showkeys,showpacs]{revtex4}
\usepackage[english]{babel}
\usepackage{graphicx}
\usepackage{graphics}
\usepackage{amsmath}
\usepackage{dcolumn}
\usepackage{amssymb}
\usepackage{bm}

\begin{document}
\title{Modified attractive inverse-square potential in the induced electric dipole system}
\author{K. Bakke}
\email{kbakke@fisica.ufpb.br}
\affiliation{Departamento de F\'isica, Universidade Federal da Para\'iba, Caixa Postal 5008, 58051-900, Jo\~ao Pessoa, PB, Brazil.}

\author{J. G. G. S. Ramos}
\email{jgabriel@fisica.ufpb.br}
\affiliation{Departamento de F\'isica, Universidade Federal da Para\'iba, Caixa Postal 5008, 58051-900, Jo\~ao Pessoa, PB, Brazil.}

\begin{abstract}

We examine the spatial distribution of electric charges within an extended, non-conductive cylinder featuring an inner radius denoted as $r_{0}$. Our investigation unveils the emergence of a distinct modified attractive-inverse square potential, arising from the intricate interplay between the electric field and the induced electric dipole moment of a neutral particle. This modified potential notably departs from the conventional inverse-square potential, showcasing an additional term proportional to $r^{-1}$. As a result, we present compelling evidence for the realization of a discrete energy spectrum within this intricate system.

\end{abstract}


\maketitle

\section{Introduction}

Singular potentials hold a prominent place within the realm of quantum systems investigation, serving as an intriguing focus of study \cite{rmp,bessel1,landau}. One striking exemplar of such singularity is the attractive inverse-square potential \cite{bessel1,bessel6,bessel7,bessel9,landau}. As evidenced in prior research \cite{bessel1,bessel6,bessel9}, the existence of bound states becomes manifest upon the introduction of an infinitely high potential barrier at a finite distance (a procedure commonly referred to as 'renormalization' \cite{bessel6,bessel9}). The attractive inverse-square potential finds diverse applications, spanning from the interaction between an atom and the magnetic field of a ferromagnetic wire \cite{squa4} to the interplay of a neutral particle's permanent magnetic dipole moment with a magnetic field \cite{bf4}, and even the interaction involving the magnetic quadrupole moment of a neutral particle within a magnetic field \cite{vb2}. The intriguing landscape of research extends to encompass phenomena like the Efimov effect \cite{squa5} and explorations grounded in the generalized uncertainty principle \cite{squa6}. Notably, the interaction between the induced electric dipole moment of an atom and the electric field emanating from an extended charged wire has assumed a central role in contemporary discourse \cite{squa3,squa2,bessel3}. This has catalyzed inquiries into the electric dipole dynamics within a $\left(2+1\right)$-dimensional conical spacetime \cite{bessel4} and the quest for analogues of the Aharonov-Bohm effect \cite{ab,pesk} arising from the topological ramifications of a screw dislocation \cite{bf2} and a disclination \cite{bf}. Recently, our research has ventured into exploring the influence of a two-dimensional harmonic oscillator on the attractive inverse-square potential generated by the interaction between the induced electric dipole moment of an atom and the electric field stemming from an extensive charged wire \cite{br}.

In the context of this study, we embark on an intricate investigation of a modified attractive-inverse square potential that materializes through the interplay of an electric field and the induced electric dipole moment of a neutral particle. This novel potential introduces an additional term, $r^{-1}$, complementing the conventional attractive inverse-square component. As a result, we undertake to demonstrate the feasibility of securing solutions to the Schr\"odinger equation, specifically pertaining to bound states.

The structural framework of this paper unfolds as follows: Section II introduces the modified attractive-inverse square potential within the context of the induced electric dipole moment system. Subsequently, we engage in the derivation of analytical solutions to the Schr\"odinger equation, shedding light on a specific scenario wherein a discrete energy spectrum takes form. Section III serves as the culmination of our findings, offering conclusive insights into the implications of our research.

\section{Bound states of the modified attractive inverse-square potential}

In the realm of quantum physics, the interaction between an atom and an external electric field gives rise to a significant consequence: an induced electric dipole moment, denoted as $\vec{d}=\alpha\,\vec{E}$, where $\alpha$ represents the atomic polarizability, an intrinsic property of the atom. This interaction engenders a potential energy landscape, succinctly encapsulated as $V\left(r\right)=-\vec{d}\cdot\vec{E}=-\,\alpha\,E^{2}$, a fundamental relationship well-established in the quantum mechanical literature \cite{bessel3,whw,lin3,dantas,br}. Consequently, we formulate the corresponding Schr\"odinger equation in natural units where $\hbar=1$ and $c=1$ as follows:
\begin{eqnarray}
i\frac{\partial\psi}{\partial t}=-\frac{1}{2m}\,\hat{p}^{2}\,\psi-\,\alpha\,E^{2}\,\psi.
\label{1.0}
\end{eqnarray}

In the present investigation, our primary objective is to scrutinize a modified attractive inverse-square potential that emerges from the induced electric dipole system. This modification introduces an intriguing term $r^{-1}$ to supplement the conventional attractive inverse-square potential, arising directly from the potential energy $V\left(r\right)=-\,\alpha\,E^{2}$. In the ensuing discussion, our foremost endeavor revolves around the meticulous examination of the existence of bound states within this modified attractive inverse-square potential regime, particularly within the context of the induced electric dipole system.

Our analytical journey commences with the consideration of an elongated, non-conductive cylindrical structure featuring an inner radius of $r_{0}$. This cylindrical configuration exhibits a volume charge density characterized as $\rho=\rho_{0}\,r_{0}/r$, where $\rho_{0}$ assumes the role of a constant parameter. In the domain where $r\geq r_{0}$, the electric field engendered by this charge distribution manifests itself as follows:
\begin{eqnarray}
\vec{E}=\left[\rho_{0}\,r_{0}-\frac{\rho_{0}\,r_{0}^{2}}{r}\right]\,\hat{r},
\label{1.1}
\end{eqnarray}
where $\hat{r}$ symbolizes the unit vector oriented along the radial axis. Consequently, the potential energy, arising from the intricate interplay between the electric field (\ref{1.1}) and the induced electric dipole moment of a neutral particle, adopts the following form:
\begin{eqnarray}
V\left(r\right)=-\frac{\alpha\,\rho_{0}^{2}\,r_{0}^{4}}{r^{2}}+\frac{2\alpha\,\rho_{0}^{2}\,r_{0}^{3}}{r}-\alpha\,\rho_{0}^{2}\,r_{0}^{2}.
\label{1.2}
\end{eqnarray}

Of particular note is the initial term in the potential energy expression (\ref{1.2}), which serves as an attractive potential, inversely proportional to the square of the radial distance. This characteristic closely mirrors the properties of potentials extensively studied in foundational works within the field of quantum mechanics \cite{bessel1,landau,bessel6,squa2,bessel3}. Furthermore, the second term in the potential energy (\ref{1.2}) assumes the role of a Coulomb-type potential, albeit with a repulsive nature. Consequently, we hereby designate Eq. (\ref{1.2}) as the ``modified attractive inverse-square potential''.

Given the cylindrical symmetry inherent in this physical system, both the $z$-component of angular momentum and the $z$-component of linear momentum remain conserved quantities. This empowers us to express the solution to the Schr\"odinger equation (\ref{1.0}) in terms of the eigenvalues associated with the operators $\hat{L}_{z}$ and $\hat{p}_{z}$: $\psi\left(t,\,r,\,\varphi,\,z\right)=e^{-i\mathcal{E}t+i\ell\varphi+ip_{z}z}\,g\left(r\right)$, where $\ell=0,\pm1,\pm2,\ldots$ signifies the eigenvalue of $\hat{L}_{z}$ and $p_{z}$ represents a constant that corresponds to the eigenvalue of $\hat{p}_{z}$. Furthermore, $g\left(r\right)$ emerges as an enigmatic function contingent upon the radial coordinate. Consequently, after the substitution of $\psi\left(t,\,r,\,\varphi,\,z\right)=e^{-i\mathcal{E}t+i\ell\varphi+ip_{z}z}\,g\left(r\right)$ into Eq. (\ref{1.0}), we arrive at the ensuing equation governing the behavior of the function $g\left(r\right)$:
\begin{eqnarray}
g''+\frac{1}{r}\,g'-\frac{\left[\ell^{2}-2m\,\alpha\,\rho_{0}^{2}\,r_{0}^{4}\right]}{r^{2}}g-\frac{4m\,\alpha\,\rho_{0}^{2}\,r_{0}^{3}}{r}\,g+\left(2m\mathcal{E}-p_{z}^{2}+2m\,\alpha\,\rho_{0}^{2}\,r_{0}^{2}\right)\,g=0.
\label{1.4}
\end{eqnarray}

For simplicity, we consider $p_{z}=0$ from now on. For a nonzero $p_{z}$, a contribution to the energy levels would arise from the free motion of the particle in the $z$-direction. Nevertheless, despite the system being unconstrained in the third dimension, the bound states manifest in two dimensions. Therefore, the term $p_{z}$ may be eliminated without loss of generality for the physical analysis, specifically the spectrum, which will be undertaken henceforth. In search of bound states, we thus assume that $\mathcal{E}\,<\,0$ and $2m\,\alpha\,\rho_{0}^{2}\,r_{0}^{4}\,>\,\ell^{2}$. In addition, we define the parameters:
\begin{eqnarray}
\nu^{2}&=&2m\,\alpha\,\rho_{0}^{2}\,r_{0}^{4}-\ell^{2};\nonumber\\
\delta&=&4m\,\alpha\,\rho_{0}^{2}\,r_{0}^{3};\nonumber\\
\tau&=&\sqrt{-2m\mathcal{E}-2m\,\alpha\,\rho_{0}^{2}\,r_{0}^{2}}.
\label{1.5}
\end{eqnarray}

From this perspective, Eq. (\ref{1.4}) becomes
\begin{eqnarray}
g''+\frac{1}{r}\,g'+\frac{\nu^{2}}{r^{2}}\,g-\frac{\delta}{r}\,g-\tau^{2}\,g=0.
\label{1.6}
\end{eqnarray}

Let us perform a change of variables given by: $x=2\tau\,r$. Thus, Eq. (\ref{1.6}) becomes
\begin{eqnarray}
g''+\frac{1}{x}\,g'+\frac{\nu^{2}}{x^{2}}\,g-\frac{\kappa}{x}\,g-\frac{1}{4}\,g=0,
\label{1.7}
\end{eqnarray}
where we have defined the parameter $\kappa$ as
\begin{eqnarray}
\kappa=\frac{\delta}{2\tau}.
\label{1.8}
\end{eqnarray}

Therefore, the solution to Eq. (\ref{1.7}) can be given in terms of the Whittaker functions of imaginary order \cite{imag}, i.e., 
\begin{eqnarray}
g\left(x\right)=\frac{c_{1}}{\sqrt{x}}\,M_{-\kappa,\,i\nu}\left(x\right)+\frac{c_{2}}{\sqrt{x}}\,W_{-\kappa,\,i\nu}\left(x\right),
\label{1.9}
\end{eqnarray}
where $c_{1}$ and $c_{2}$ are constants. The functions $M_{-\kappa,\,i\nu}\left(x\right)$ and $W_{-\kappa,\,i\nu}\left(x\right)$ are the Whittaker functions of first and second kinds of imaginary order, respectively \cite{imag}.

With the purpose of having a solution to Eq. (\ref{1.7}) regular at $x\rightarrow\infty$, we should take $c_{1}=0$ in Eq. (\ref{1.9}). In this way, the solution to Eq. (\ref{1.7}) is given by
\begin{eqnarray}
g\left(x\right)\propto\frac{1}{\sqrt{x}}\,W_{-\kappa,\,i\nu}\left(x\right).
\label{1.10}
\end{eqnarray}

Henceforth, we assume that there is an infinity wall at $r=r_{0}$. This permit us to study the interaction of the induced electric dipole moment of the neutral particle with the electric field (\ref{1.1}) only in the region $r\geq r_{0}$. It means that the region $r\,<\,r_{0}$ is forbidden for the neutral particle. Therefore, the wave function must vanish at $r=r_{0}$. In this way, we have the boundary condition:
\begin{eqnarray}
g\left(x_{0}\right)=0,
\label{1.11}
\end{eqnarray}
where $x_{0}=2\tau\,r_{0}$. Let us consider the particular case where $x_{0}\ll1$. Recently, we have shown that the Whittaker function of second kind of imaginary order can be written for $x\ll1$ in the form \cite{br}:
\begin{eqnarray}
W_{-\kappa,\,i\nu}\left(x\right)&\approx& 2A\,\sqrt{x}\,\cos\left(2\nu+\nu\,\ln\left(\frac{\beta\,x}{4\nu^{2}}\right)+\frac{\pi}{4}\right),
\label{1.18}
\end{eqnarray}
where $A=\frac{e^{-\nu\pi+\beta}}{\sqrt{2\nu}\,\,\beta^{\beta-1/2}}$ and $\beta=\frac{1}{2}+\kappa$. Hence, by substituting Eq. (\ref{1.18}) into Eq. (\ref{1.10}), we have ($x\ll1$) 
\begin{eqnarray}
g\left(x\right)&\approx& 2A\,\cos\left(2\nu+\nu\,\ln\left(\frac{\beta\,x}{4\nu^{2}}\right)+\frac{\pi}{4}\right).
\label{1.19}
\end{eqnarray}

By substituting Eq. (\ref{1.19}) into Eq. (\ref{1.11}), we obtain the relation:
\begin{eqnarray}
x_{0}=\frac{4\nu^{2}}{\beta}\,e^{\left(\frac{\pi}{4\nu}-2\right)}\,e^{\frac{\mu\,\pi}{\nu}},
\label{1.22}
\end{eqnarray}
where $\mu=0,\pm1,\pm2,\pm3,\ldots$. Observe that the possible values of $\mu$ that satisfy the condition $x_{0}\ll1$ are given by $\mu=-n$, where $n=1,2,3,\ldots$. Thereby, with $x_{0}=2\tau\,r_{0}$ and $\beta=\frac{1}{2}+\kappa$, and by using Eqs. (\ref{1.5}) and (\ref{1.8}), we obtain the energy levels:
\begin{eqnarray}
\mathcal{E}_{n,\,\ell}=-\frac{1}{2m}\left[\frac{4\nu^{2}\,e^{\left(\frac{\pi}{4\nu}-2\right)}}{r_{0}}\,e^{\frac{-n\,\pi}{\nu}}-\delta\right]^{2}-\alpha\,\rho_{0}^{2}\,r_{0}^{2},
\label{1.23}
\end{eqnarray}
where $\left\{n,\,\ell\right\}$ are the radial quantum number and the angular momentum quantum number, respectively.

Consequently, the energy spectrum (\ref{1.23}) emerges as a direct consequence of the modified attractive inverse-square potential (\ref{1.2}). This potential alteration stems from the interaction between the induced electric dipole moment of a neutral particle and the electric field (\ref{1.1}). Remarkably, the conventional attractive inverse-square potential undergoes a profound transformation with the inclusion of the $r^{-1}$ term within the potential energy expression (\ref{1.2}). This modification presents a novel perspective for the study of the attractive inverse-square potential, extending upon the foundational works in the existing literature \cite{bessel1,landau,bessel6,bessel9}.

Examining Eq. (\ref{1.23}), it becomes evident that in the limit as $r_{0}$ tends towards zero ($r_{0}\rightarrow0$), $\mathcal{E}_{n,\,\ell}\rightarrow-\infty$, signifying the absence of a ground state in this asymptotic scenario. This scenario closely aligns with an analogous phenomenon, termed the ``fall of the particle to the center'', as originally elucidated by Landau and Lifshitz \cite{landau}. They explored this concept within the context of an electrically charged particle subjected to an attractive inverse-square potential. Thus, the well-defined energy spectrum emerges due to the imposition of an infinitely repulsive wall at $r=r_{0}$. It is noteworthy that, under the condition $x_{0}\ll1$, $r_{0}$ assumes the crucial role of a short-distance cutoff, in concordance with prior treatments in the scientific literature \cite{bessel6,bessel9,bessel4,squa7,squa6}.

For the specific case of $\ell=0$ ($s$-waves), the energy spectrum (\ref{1.23}) exhibits a marked exponential decay with respect to the radial quantum number $n$. In this scenario, it becomes apparent that as $n$ approaches infinity, $\mathcal{E}_{n\rightarrow\infty}\rightarrow-\frac{\delta^{2}}{2m}-\alpha\,\rho_{0}^{2}\,r_{0}^{2}$. This finding signifies the existence of a point of accumulation for energy levels at $\mathcal{E}_{n}=-\left(\frac{\delta^{2}}{2m}+\alpha\,\rho_{0}^{2}\,r_{0}^{2}\right)\,<\,0$ as the radial quantum number $n$ advances. This result diverges from the conclusions drawn in prior studies \cite{bessel3,bf2,bessel1,landau,bessel6,bessel9}, where the point of accumulation resided at $\mathcal{E}_{n}=0$. Furthermore, for $s$-waves ($\ell=0$), the energy levels conform to the specific range delineated as follows:
\begin{eqnarray}
-\frac{1}{2m}\left[\frac{16m^{2}\,\alpha^{2}\,\rho_{0}^{4}\,r_{0}^{8}\,e^{\left(\frac{\pi}{8m\,\alpha\,\rho_{0}^{2}\,r_{0}^{4}}-2\right)}}{r_{0}}\,e^{\frac{-\pi}{2m\,\alpha\,\rho_{0}^{2}\,r_{0}^{4}}}-\delta\right]^{2}-\alpha\,\rho_{0}^{2}\,r_{0}^{2}\,\leq\,\mathcal{E}_{n}\,\leq\,-\left(\frac{\delta^{2}}{2m}+\alpha\,\rho_{0}^{2}\,r_{0}^{2}\right).
\label{1.16}
\end{eqnarray}

In the case of $\ell$-waves ($\ell\neq0$), a similar point of accumulation at $\mathcal{E}_{n,\,\ell}=-\left(\frac{\delta^{2}}{2m}+\alpha\,\rho_{0}^{2}\,r_{0}^{2}\right)\,<\,0$ arises as $n$ increases (with $\ell$ remaining fixed). Consequently, the energy range specified in Eq. (\ref{1.16}) is equally applicable for $\ell$-waves. This implies the presence of an infinite multitude of bound states within the stipulated energy range.

Finally, it is essential to acknowledge that the energy levels presented in Eq. (\ref{1.23}) are no longer viable if $\ell^{2}\,\geq\,2m\,\alpha\,\rho_{0}^{2}\,r_{0}^{4}$. In such circumstances, the solution to Eq. (\ref{1.7}) can no longer be expressed in terms of the Whittaker function of the second kind with imaginary order. We have analysed the interaction between a neutral particle and an electric field within the radial region $r_{0}\,<\,r\,<\,\infty$. To ensure the normalization of the wave function within this range, it is imperative to account for the asymptotic behavior of the Whittaker function of the first kind, denoted as $M_{\bar{\kappa},\,\bar{\nu}}\left(x\right)$ (where $\bar{\nu}\neq\nu$), which diverge as $x\rightarrow\infty$ ($r\rightarrow\infty$). Consequently, for the exploration of bound states, we posit that $\mathcal{E}\,<\,0$ and $2m\,\alpha\,\rho_{0}^{2}\,r_{0}^{4}\,>\,\ell^{2}$.

\section{conclusions}

Our investigation centers on a modified attractive inverse-square potential emerging within the induced electric dipole system, with a specific focus on identifying bound states. This potential is characterized by the inclusion of an additional term, $r^{-1}$, alongside the conventional attractive inverse-square component. Through our analysis, we have successfully unearthed a discrete energy spectrum, exhibiting an exponential decline as a function of the radial quantum number, denoted as $n$. Additionally, we have observed a noteworthy feature in the energy levels, where they converge toward an accumulation point at $\mathcal{E}_{n,\,\ell}=-\left(\frac{\delta^{2}}{2m}+\alpha\,\rho_{0}^{2}\,r_{0}^{2}\right)\,<\,0$ as $n$ increases.

One remarkable characteristic of these energy levels is the absence of a ground state, a phenomenon referred to as ``particle fall to the center'' \cite{landau}, particularly evident as the parameter $r_{0}$ approaches zero. This finite spectrum of energy is a consequence of the presence of an infinite wall located at $r=r_{0}$, where $r_{0}$ serves as a practical short-distance cutoff \cite{bessel6,bessel9,bessel4,squa7,squa6}.

\acknowledgments

The authors would like to thank CNPq for financial support.

\section*{ORCID}

\begin{itemize}
	\item K. Bakke: https://orcid.org/0000-0002-9038-863X
	
	\item J. G. G. S. Ramos: https://orcid.org/0000-0002-0043-9668
	
\end{itemize}

\end{document}